\documentclass[format=sigconf, nonacm]{acmart}

\usepackage{xspace}
\usepackage{amsmath}
\usepackage{amssymb}
\usepackage{paralist}
\usepackage{verbatim}
\usepackage{url}
\usepackage{xytree}
\usepackage[all]{xy}
\usepackage{multirow}
\usepackage{graphicx}
\usepackage{subcaption}

\newcommand{\Dfn}[1]{\textbf{\emph{#1}}}

\newcommand{\Nat}{\ensuremath{\mathbb{N}}\xspace}
\newcommand{\Powerset}[1]{\ensuremath{2^{#1}}}

\newcommand{\anEA}{\ensuremath{M}\xspace}
\newcommand{\EA}[4]{\ensuremath{\langle #1, #2, #3, #4 \rangle}}
\newcommand{\EASymbols}{\ensuremath{\Sigma}\xspace}
\newcommand{\EAStates}{\ensuremath{Q}\xspace}
\newcommand{\EAInit}{\ensuremath{q_0}\xspace}
\newcommand{\EATrans}{\ensuremath{\delta}\xspace}
\newcommand{\anEAState}{\ensuremath{q}\xspace}

\newcommand{\anEvent}{\ensuremath{a}\xspace}
\newcommand{\anotherEvent}{\ensuremath{b}\xspace}
\newcommand{\yetAnotherEvent}{\ensuremath{c}\xspace}
\newcommand{\anEventSeq}{\ensuremath{w}\xspace}
\newcommand{\anotherEventSeq}{\ensuremath{u}\xspace}
\newcommand{\yetAnotherEventSeq}{\ensuremath{v}\xspace}

\newcommand{\aSystem}{\ensuremath{\chi}\xspace}
\newcommand{\System}[5]{\ensuremath{\langle #1, #2, #3, #4, #5 \rangle}}

\newcommand{\aConGraph}{\ensuremath{\mathcal{CG}}\xspace}
\newcommand{\ConGraph}[3]{\ensuremath{\langle #1, #2, #3 \rangle}}
\newcommand{\Entities}{\ensuremath{\mathcal{E}}\xspace}
\newcommand{\Devices}{\ensuremath{\mathcal{D}}\xspace}
\newcommand{\Brokers}{\ensuremath{\mathcal{B}}\xspace}
\newcommand{\Link}{\ensuremath{\mathit{link}}\xspace}
\newcommand{\Publish}{\ensuremath{\mathit{publish}}\xspace}
\newcommand{\Notify}{\ensuremath{\mathit{notify}}\xspace}
\newcommand{\Bridge}{\ensuremath{\mathit{bridge}}\xspace}
\newcommand{\Monitored}{\ensuremath{\mathit{monitored}\xspace}}

\newcommand{\anEntity}{\ensuremath{x}\xspace}
\newcommand{\anotherEntity}{\ensuremath{y}\xspace}
\newcommand{\yetAnotherEntity}{\ensuremath{z}\xspace}

\newcommand{\anEventPolicy}{\ensuremath{\mathcal{EP}}\xspace}
\newcommand{\EventPolicy}[4]{\ensuremath{\langle #1, #2, #3, #4 \rangle}}
\newcommand{\Events}{\ensuremath{\EASymbols}\xspace}
\newcommand{\Trans}{\ensuremath{\Delta}\xspace}

\newcommand{\aBrokerPolicy}{\ensuremath{\mathcal{BP}}\xspace}
\newcommand{\BrokerPolicy}[3]{\ensuremath{\langle #1, #2, #3 \rangle}}
\newcommand{\BPTypes}{\ensuremath{\mathcal{T}}\xspace}
\newcommand{\BPTyping}{\ensuremath{\mathit{type}}\xspace}
\newcommand{\BPAllow}{\ensuremath{\mathit{allow}}\xspace}
\newcommand{\BPPropagate}{\ensuremath{\mathit{propagate}}\xspace}
\newcommand{\aBPType}{\ensuremath{t}\xspace}

\newcommand{\Sub}{\ensuremath{\mathit{sub}}\xspace}

\newcommand{\aSysState}{\ensuremath{\gamma}\xspace}
\newcommand{\theInitSysState}{\ensuremath{\gamma_{\mathit{init}}}\xspace}
\newcommand{\SysState}[3]{\ensuremath{\langle #1, #2, #3 \rangle}}
\newcommand{\aTS}{\ensuremath{t}\xspace}
\newcommand{\SSST}{\ensuremath{\mathit{ST}}\xspace}
\newcommand{\SSWL}{\ensuremath{\mathit{WL}}\xspace}
\newcommand{\SysStates}[1]{\ensuremath{\Gamma_{#1}}}

\newcommand{\TaskBroker}[3]{\ensuremath{\mathsf{broker}(#1, #2, #3)}}

\newcommand{\TaskTransmit}[3]{\ensuremath{\mathsf{transmit}(#1, #2,
    #3)}}
\newcommand{\aTask}{\ensuremath{\tau}\xspace}
\newcommand{\Tasks}[1]{\ensuremath{\mathit{TK}_{#1}}}
\newcommand{\Annotate}[3]{\ensuremath{#1[#2, #3]}}
\newcommand{\AnnTasks}[1]{\ensuremath{\mathit{AT}_{#1}}}
\newcommand{\anAnnTask}{\ensuremath{\alpha}\xspace}
\newcommand{\TaskOrder}{\ensuremath{\sqsubseteq}\xspace}
\newcommand{\TaskSelect}{\ensuremath{\mathit{select}}\xspace}

\newcommand{\aTSPub}{\ensuremath{\aTS_\mathit{pub}}\xspace}
\newcommand{\aTSGen}{\ensuremath{\aTS_\mathit{gen}}\xspace}

\newcommand{\Observations}[1]{\ensuremath{\Lambda_{#1}}}
\newcommand{\anObs}{\ensuremath{\lambda}\xspace}
\newcommand{\obsIn}[3]{\ensuremath{\mathsf{in}(#1, #2, #3)}}
\newcommand{\obsOut}[3]{\ensuremath{\mathsf{out}(#1, #2, #3)}}
\newcommand{\anObsSeq}{\ensuremath{\pi}\xspace}

\newcommand{\SysTransBinRel}[1]{\ensuremath{\rightarrow_{#1}}}
\newcommand{\SysMultiTransBinRel}[1]{\ensuremath{\Longrightarrow_{#1}}}

\newcommand{\Mapper}[1]{\ensuremath{\mathit{mapper}(#1)}}
\newcommand{\Filter}[1]{\ensuremath{\mathit{filter}(#1)}}

\newcommand{\Interleaving}[1]{\ensuremath{\mathit{inter}(#1)}}

\newcommand{\Indis}{\ensuremath{\sim}}

\newcommand{\NormQoS}[1]{\textbf{Norm-#1}}
\newcommand{\UniqueEvents}{\ensuremath{\Pi}\xspace}
\newcommand{\Signature}{\ensuremath{\sigma}\xspace}

\newcommand{\ExtractIn}[2]{\ensuremath{\mathit{IN}(#1 \mid #2)}}
\newcommand{\ExtractOut}[3]{\ensuremath{\mathit{OUT}(#1 \mid #2, #3)}}
\newcommand{\Produces}{\ensuremath{\leadsto}\xspace}

\newtheorem{convention}{Convention}

\begin{document}


\title{Brokering Policies and Execution Monitors for IoT Middleware}
\author{Juan Carlos Fuentes Carranza}
\affiliation{University of Calgary}
\email{krloswd@gmail.com}
\author{Philip W.L.~Fong}
\affiliation{University of Calgary}
\email{pwlfong@ucalgary.ca}

\begin{abstract}
  Event-based systems lie at the heart of many cloud-based
  Internet-of-Things (IoT) platforms. This combination of the Broker
  architectural style and the Publisher-Subscriber design pattern
  provides a way for smart devices to communicate and coordinate with
  one another.  The present design of these cloud-based IoT frameworks
  lacks measures to (i) protect devices against malicious cloud
  disconnections, (ii) impose information flow control among
  communicating parties, and (iii) enforce coordination protocols in
  the presence of compromised devices.  In this work, we propose to
  extend the modular event-based system architecture of Fiege \emph{et
    al.}, to incorporate brokering policies and execution monitors, in
  order to address the three protection challenges mentioned above. We
  formalized the operational semantics of our protection scheme,
  explored how the scheme can be used to enforce BLP-style information
  flow control and RBAC-style protection domains, implemented the
  proposal in an open-source MQTT broker, and evaluated the
  performance impact of the protection mechanisms.
\end{abstract}

\maketitle

\section{Introduction}

The Internet of Things (IoT) is envisioned to be a massively
distributed system of smart devices, each potentially equipped with
physical sensors and actuators.  This is, however, not the first day
we develop distributed systems.  Decades of experiences in
architecting distributed systems have gone into the design of IoT
frameworks. Specifically, the Broker architectural style has been
employed to coordinate the communication between devices, so that the
idiosyncrasies of low-level network communications can be abstracted
away \cite{buschmann}.  The Publisher-Subscriber design pattern has
been applied to prevent the need for polling the state of
collaborating devices, and to provide decoupling between message
senders and receivers \cite{buschmann}.  In this work, we use the term
\Dfn{event-based systems} to refer to this combination of Broker and
Publisher-Subscriber \cite{Fiege-etal:2002:KRR}.  Many existing IoT
frameworks are essentially variants of event-based systems.  The MQTT
protocol, a messaging transport protocol widely used in popular IoT
frameworks (including that of IBM, Amazon, and Microsoft), is an
example of such a design \cite{Mosquitto}.

The implementation of event-based systems in the context of commercial
IoT frameworks, however, has a twist, and that is the centralization
of the notification mechanism on a cloud platform.  This is
understandable as a cloud-based implementation offers the benefit of
low configuration and maintenance cost.

The motivation of this work is to inquire about what unique protection
challenges are presented in the cloud-based realization of event-based
systems, and to devise protection mechanisms and access control
technologies for addressing them. We discern three protection
challenges.
\begin{asparaenum}
\item \Dfn{Cloud disconnection.} Previous work has argued that a
  cloud-based implementation of an IoT coordination framework could be
  susceptible to \Dfn{cloud disconnection} \cite{Tam}. That is, the
  cloud could be disconnected because it has fallen victim to a DDoS
  attack, or the ISP becomes unavailable because of an Internet-scale
  incident, or the physical connection to the cloud is maliciously
  severed.
\item \Dfn{Information flow control.}  Some of the messages exchanged
  by devices are sensitive. The nature of a Publisher-Subscriber
  system is that the two parties of communication need not know one
  another.  Such decoupling must be complemented by proper information
  flow control in order that messages are not accidentally received by
  unintended parties.  This need for privacy and confidentiality is
  particularly relevant in the era of ubiquitous deployment of sensors,
  which collect personal information from health data to what goes on
  in the bedroom.
\item \Dfn{Enforcing coordination protocols.}  Smart devices are
  limited in capability, and thus they are easy targets for malicious
  exploits. Yet these devices are now part of our safety
  infrastructure. An example is to have the smoke detector
  shut down the furnace in the event of a fire.  If these devices are
  compromised, then the coordination logic that we rely on to ensure
  safety will no longer be carried out properly. It is therefore
  important to build measures into the event-based systems so that any
  deviations from coordination protocols can be detected, and, if
  possible, compensated for.
\end{asparaenum}

This work proposes a unified protection scheme that addresses the
aforementioned challenges in event-based systems.  More specifically,
the following are our contributions.
\begin{asparaenum}
\item We adopted the modular event-based system of Fiege
  \emph{et al} \cite{Fiege-etal:2002:SAC, Fiege-etal:2002:ECOOP,
    Fiege-etal:2002:KRR, Muhl-etal:2006}, which is made up of multiple
  communicating brokers, to mitigate the impact of cloud disconnection
  (\S \ref{sec-background}).  More importantly, we extended their
  work, and introduced the ideas of imposing \Dfn{brokering policies}
  to control the flow of information among brokers and devices (\S
  \ref{sec-brokering}), and interposing \Dfn{execution monitors} along
  network links to enforce coordination protocols (\S
  \ref{sec-exec-mon}).  We demonstrated the utility of our proposal in
  a case study that shows how the proposed protection mechanisms can
  be deployed to protect a smart home (\S \ref{sec-case-study}).
\item We formalized the operational semantics of our protection scheme
  in the form of a state transition model (\S \ref{sec-model}). This
  lays the foundation for the study of policy enforceability in the
  context of event-based systems. 
\item We devised a number of usage patterns that allow one to impose
  BLP-style information flow control and RBAC-style protection domains
  in the context of an event-based system (\S \ref{sec-usage}).
\item We implemented the proposed protection scheme in Mosquitto, an
  open-source MQTT broker (\S \ref{sec-implementation}), and evaluated
  the performance impact of the proposed protection mechanisms (\S
  \ref{sec-performance}).
\end{asparaenum}

\section{Related Work}

This work builds on the modular event-based system architecture of
Fiege \emph{et al.} \cite{Fiege-etal:2002:SAC, Fiege-etal:2002:ECOOP,
  Fiege-etal:2002:KRR, Muhl-etal:2006}, but moves significantly beyond
it.  The adoption of multiple brokers is no longer motivated by
modularity, but for anticipating cloud disconnection. Brokering
policies are generalization of the Fiege \emph{et al.} scoping rule
(\S \ref{sec-background}). Execution monitors are
generalization of event filters and mappers (\S \ref{sec-acc-con-EA}).

Cloud disconnection is addressed in \cite{Tam} by having a local hub
that replicates the coordination scripts that are stored in the cloud.
Our work can be seen as a more principled and general way to localize
coordination logics. Virtual objects are software shadows of physical
devices \cite{Alshehri-Sandhu:2016, Alshehri-Sandhu:2017}. Having
virtual objects hosted in the cloud partly addresses the problem of
occasional cloud disconnections, but this does not guarantee the
proper functioning of coordination logics when the cloud is
disconnected. Pushing security enforcement towards the edge of the
network is a general trend \cite{Sicari-etal:2017,
  Phung-etal:2017}. Our work provides a general architecture for
hosting safety-related coordination logics at the edge in anticipation
of cloud disconnection.

There has been growing interests in applying Attribute-Based Access
Control (ABAC) \cite{ABAC} in the context of IoT in general, and
event-based systems in particular \cite{Sciancalepore-etal:2016,
  Rizzardi-etal:2016, Gabillon-Bruno:2018, Colombo-Ferrari:2018}.  Our
work is complementary. For example, ABAC can be deployed to control
which devices can be connected to the various brokers in a broker
ensemble.  Each broker plays the role of a protection domain.
Information flow control (\S \ref{sec-BLP}) and history-tracking
execution monitors (\S \ref{sec-exec-mon}) can then be employed to
control the routing and transformation of events among protection
domains (brokers). This is the setup envisioned in \S \ref{sec-RBAC}.
Note that when we use the RBAC terminology of \cite{rbac-middleware}
in \S \ref{sec-RBAC}, we are not sanctioning RBAC for IoT. The so-called
``roles'' are but protection domains.

Also relevant to this work are attempts to add content-based access
control to MQTT \cite{Colombo-Ferrari:2018}. A technical challenge is
to meet the hungry throughput demand of IoT applications. Our
implementation of history-tracking execution monitors is a first step
towards realizing high-throughput, content-based event processing in a
messaging system (\S \ref{sec-implementation}--\ref{sec-performance}).

\section{Background: Multiple Brokers}
\label{sec-background}

\paragraph{Distributed Event-Based Systems.} 
A popular system architecture for IoT applications is to have devices
connected to a cloud-based broker.  Typically a device signals a
change of its state by publishing a message, also known as a
\Dfn{event}, through the broker.  Each event is typically annotated
with an \Dfn{event topic}, which is a tag that allows the broker to
quickly classify events for the purpose of message delivery.  Devices
may also declare to the broker what event topics they are interested
in.  This declaration of interest is also known as a
\Dfn{subscription}. On receiving an event published under a certain
event topic, the broker will notify all the devices that have
previously subscribed to that event topic.  The rationale for adopting
such an architecture in a distributed system is so that devices are
decoupled from one another: the publisher does not need to know who
will end up receiving a message, and a subscriber does not need to
know from where the event originates. The broker is the only one who
needs to know the network addresses of the communicating parties.  In
addition, subscription-based notification eliminates the need for
periodic polling.  This architectural design, in which the \Dfn{broker
  architecture style} is combined with the \Dfn{publisher-subscriber
  design pattern} \cite{buschmann}, is known in the literature as an
\Dfn{event-based system} \cite{Fiege-etal:2002:KRR}.
MQTT, a popular messaging transport protocol designed for IoT
applications, is a typical instantiation of this system architecture
\cite{MQTT}.

\paragraph{Cloud Disconnection.} 
Previous work has argued that an IoT framework that is built around a
centralized, cloud-based broker is susceptible to \Dfn{cloud
  disconnection} \cite{Tam}.  The latter could be caused by, for
example, the cloud infrastructure becoming the victim of a DDoS
attack, the network infrastructure of an Internet Service Provider
(ISP) being affected by an internet outage event, or a malicious party
physically severing the connection between the devices and the
cloud-based broker.  Such disconnections could have physical safety
implications.  For example, the cloud-based broker typically hosts
\Dfn{coordination logics} that notify one device when the state of
another device changes.  Such coordination logics could very well be
part of the measures for ensuring physical safety: e.g., notify
homeowner in case of break-in, shutdown furnace in case smoke is
detected, unlock door for helper in case an elderly occupant is found
to have fallen on the ground, etc.  If the cloud is maliciously
disconnected, then the coordination logics hosted on the cloud will not
be executed. This could have potentially serious safety implications.

\paragraph{Enter Multiple Brokers.}

To anticipate cloud disconnection, previous work has argued that
redundancy must be built into an IoT framework \cite{Tam}.  In other
words, it is desirable to ensure that coordination logics are still
executed in the event of cloud disconnection.

The modular event-based system architecture of Fiege \emph{et al.}
\cite{Fiege-etal:2002:SAC, Fiege-etal:2002:ECOOP, Fiege-etal:2002:KRR,
  Muhl-etal:2006} provides a good starting point for anticipating
cloud disconnection.  The system design of Fiege was originally
proposed for the sake of modularizing distributed event-based systems,
in this work we reappropriate their design for building resiliency
into an IoT framework so as to mitigate the impact of cloud
disconnection.  Ultimately, we will move beyond their proposal (see \S
\ref{sec-overview}).

\begin{figure}
  \centering
\mbox{%
\xymatrix@R=8pt@C=10pt{
  & *+[o][F]{A} \ar@{-}[dl] \ar@{-}[dr]
  & &
  *+[o][F]{B} \ar@{-}[dl] \ar@{-}[dr] & \\
  *+[o][F]{C} \ar@{-}[d] \ar@{-}[dr] 
  & & *+[o][F]{D} \ar@{-}[d]
  & & *+[o][F]{E} \ar@{-}[dl] \ar@{-}[d]
  \\
  *+[F]{1} & *+[F]{2} & *+[F]{3} & *+[F]{4} & *+[F]{5} 
}}
\caption{A sample hierarchy of brokers and devices. Ovals represent
  brokers, while boxes represent devices.\label{fig-hierarchy}}
\end{figure}
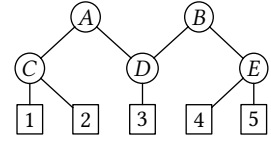

In the system architecture of Fiege \emph{et al.}, there are multiple
brokers connected to one another through network connections.  When a
device publishes an event to a broker, the latter does not only notify
the subscriber devices who are connected directly to itself, but also
passes the event to its neighbouring brokers.  This propagation of
message publication is further regulated by two mechanisms.
\begin{asparaenum}
\item The design goal of Fiege \emph{et al.} is to provide modularity
  in the construction of distributed systems. Consequently, they
  envisioned that each broker is the physical embodiment of a
  namespace of events.  Recall that namespace management is the
  classical means for adding modularity to a programming language
  environment.  Influenced by this programming language root, Fiege
  \emph{el al.} called these namespaces \Dfn{scopes}.  To support
  nested scoping (as in a structured programming language), they did
  two things. First, they organized the brokers into a hierarchy.
  Second, they imposed a scoping rule on the propagation of events
  from one broker to another.
  \begin{description}
  \item[The Fiege \emph{et al.} Scoping Rule.]  An event published in
    a broker \anEntity is visible to the subscribers in broker
    \anotherEntity only if there exists a broker \yetAnotherEntity
    that is a common ancestor of both \anEntity and \anotherEntity in
    the broker hierarchy.  A degenerate case is when \yetAnotherEntity
    is either \anEntity or \anotherEntity.
  \end{description}
  For example, in Fig.~\ref{fig-hierarchy}, an event published by
  device $1$ to broker $C$ will be visible to subscriber $2$ under
  broker $C$ and also subscriber $3$ under broker $D$. The latter case
  is justified by $A$ being a common ancestor of $C$ and $D$.  On the
  contrary, subscribers under broker $E$ will not be notified, as $E$
  does not share a common ancestor with $C$.

  The scoping rule above can be implemented by a simple modification
  to the broker. In particular, if an event is propagated to a broker
  \anEntity from one of its parents, then \anEntity will not propagate
  the event to its other parents.  For example, in
  Fig.~\ref{fig-hierarchy}, broker $D$ will not propagate to parent
  broker $B$ those events received from parent broker $A$.
\item To push the scoping metaphor further, Fiege \emph{et al.} also
  introduced \Dfn{event filtering} and \Dfn{event mapping} into the
  event propagation mechanism.  Event filtering refers to a mechanism
  that allows the system architect to specify that events of certain
  topics are not permitted to be propagated along a network connection
  linking one broker to another.  This promotes \Dfn{encapsulation},
  and thus allows certain event topics to become private within a
  scope (i.e., a broker).  Event mapping refers to a mechanism that
  allows the system architect to specify how event topics are ``renamed''
  (to other topics) as they pass through a network connection
  linking one broker to another.  This further promotes encapsulation.
\end{asparaenum}

The introduction of multiple brokers offers a good starting point for
anticipating cloud disconnection. For example, in
Fig.~\ref{fig-hierarchy}, if broker $A$ is disconnected, then other
brokers can still perform local propagation of events. We envision
that brokers are organized based on their network proximity to the
devices.  For example, a local hub can connect devices belonging to
the same local area network. Then an ISP may host a broker for each
geographical region it serves.  Such a broker will be the parent of
the local hubs.  A cloud service provider then provides a top-level
broker, with ISP brokers as children. Other intermediate level brokers
can be introduced as needed (e.g., brokers mirroring organizational
divisions).  Coordination logics hosted at one level of the hierarchy
will not be affected by the disconnection of brokers at a higher
level.

\section{Overview of Protection Scheme}
\label{sec-overview}

While the introduction of a broker hierarchy does serve as a good
starting point for protecting an IoT framework against malicious or
accidental cloud disconnection, it is the position of this work that
the scoping rule and event filtering/mapping of Fiege \emph{et al.} do
not go far enough for addressing the access control needs of an IoT
framework, we therefore propose two additional protection features in
this work.
\begin{asparaenum}
\item \emph{Controlling flow of information.}  While encapsulation
  allows the system architect to better reason about the behavior of
  individual distributed software components and compose them together
  with ease, it does not allow the \emph{security architect} to impose
  finer-grained control of information flow within the broker
  ensemble.  Certain brokers may host devices that publish sensitive
  messages (e.g., a broker that are connected to sensors in the master
  bedroom of a smart home), and thus the security architect may want
  to make sure that such events are received only by a selected group
  of authorized devices.  One way to achieve this is to control the
  information flow paths in the broker ensemble.  To this end we
  propose a generalization of Fiege \emph{et al.}'s architecture, so that
  arbitrary brokering policies can be imposed on the broker ensemble
  to control information flow paths. The scoping rule of Fiege
  \emph{et al.} becomes a special case of these brokering policies.
\item \emph{Enforcing collaboration protocols.}  When devices
  collaborate with one another, there are typically well-understood
  collaboration protocols that they have to follow.  Unfortunately,
  IoT devices have limited capabilities, and thus they are easy to
  compromise. In fact, we should take as an axiom that they will be
  compromised sooner or later. Compromised devices may not follow the
  intended collaboration protocol, and thus the integrity of the
  collaboration will be destroyed.  We propose an extension of Fiege
  \emph{et al.}'s architecture, so that execution monitors can be put
  in place to detect violation of the collaboration protocol, or even
  enforce the protocol by modifying the message exchanged by the
  devices.  The event filtering and event mapping features of Fiege
  \emph{et al.}'s architecture become special cases of execution
  monitors.
\end{asparaenum}

\subsection{Configurable Brokering Policies}
\label{sec-brokering}

Rather than having a fixed scoping rule to regulate how events are
propagated among brokers, our scheme allows the security architect to
impose arbitrary \Dfn{brokering policies}.  A brokering policy is a
system-wide scheme for deciding how each broker is to forward events.
The broker ensemble is no longer organized as a hierarchy. Instead,
brokers are peers that propagate events in accordance to the brokering
policies.  A brokering policy is composed of the following elements.

\begin{asparaenum}
\item When two brokers (or a device and a broker) are connected to one
  another, we treat each direction of communication a distinct
  \Dfn{network
  link}.  Each network link is assigned a \Dfn{link type}.  These link
  types are domain specific, and the security architect can choose to
  create link types that are relevant to the application domain in
  question. For example, when we model Bell-LaPadula (BLP)-style
  information flow control, a link type could be
  ``\textsf{confidential}'', indicating that the events passing
  through the network links with this label carry confidential
  information (\S \ref{sec-BLP}).  Alternatively, to model the Fiege
  \emph{et al.} broker hierarchy, a network link from a child broker
  to a parent broker can be associated with the link type
  ``\textsf{up}'' (\S \ref{sec-component}).
\item A \Dfn{brokering table}, which is replicated to all the brokers,
  specifies if a broker may forward an event that it receives from a
  network link of type $\aBPType_1$ to a network link of type
  $\aBPType_2$.  In short, the brokering table is an $n \times n$
  table of boolean values, where $n$ is the number of link types.
\end{asparaenum}

As we shall see, the provision for configurable brokering policies
allow the security architect to control the information flow paths
in the broker ensemble (\S \ref{sec-BLP}).

\subsection{History-Tracking Execution Monitors}
\label{sec-exec-mon}

Rather than imposing an event filter or an event mapper along network
links, our scheme allows arbitrary execution monitors to be interposed
along network links.  An execution monitor intercepts every event
moving along a network link, and responds by (a) suppressing the
event, (b) leaving the event as is, or (c) transforming the event to
another sequence of events. Note that option (c) may involve injecting
new events into the network link.  More importantly, an execution
monitor is stateful, thereby enabling the execution monitor to enforce
History-Based Access Control (HBAC) policies \cite{Evans-Twyman:1999,
  Schneider:2000, IRM, Wallach-etal:2000, Fong:2004, Krukow-etal:2008,
  EA:IJIS, EA:TISSEC}.  This history-tracking feature of an execution
monitor is the key to enforcing collaboration protocols among devices
(\S \ref{sec-case-study}).  We will demonstrate that event filtering
and event mapping are but special cases for our execution monitors (\S
\ref{sec-acc-con-EA}).

Let us make these notions formal.  Let \Events be an alphabet of
\Dfn{events}.  (More precisely, these are \Dfn{event topics}. We
identify events by their event topics when the event payloads are not
our focus.)  We use \anEvent, \anotherEvent and \yetAnotherEvent to
denote members of \Events, and write \anotherEventSeq,
\yetAnotherEventSeq and \anEventSeq to denote members of $\Events^*$.

Ligatti \emph{et al.} proposed the \Dfn{edit automaton (EA)} as an
abstract model of execution monitors. An edit automaton transforms a
sequence of events generated by an untrusted program to another
sequence of events that is safe for the observer to consume
\cite{EA:IJIS, EA:TISSEC}.  The idea is that certain event sequences
may damage the integrity of the observer, and thus to protect the
observer those sequences are ``rewritten'' to a benign form.  We have
adapted their definition to a form that is readily implementable.
More specifically, we define an \Dfn{edit automaton (EA)} to be a
quadruple \EA{ \EASymbols }{ \EAStates }{ \EAInit }{ \EATrans }, such
that \EASymbols is a finite set of symbols, \EAStates is a countable
set of states, and $\EAInit \in \EAStates$ is the initial state, and
$\EATrans : \EAStates \times \EASymbols \rightarrow \EAStates \times
\EASymbols^*$ is the transition function (NB: a total function).
Given a current state \anEAState and an input symbol \anEvent,
$\EATrans (\anEAState, \anEvent)$ is a pair
$(\anEAState', \anEventSeq)$, where $\anEAState'$ is the next state,
and \anEventSeq is a sequence of output symbols generated by the
transition. If the output sequence \anEventSeq is empty (i.e.,
$\epsilon$), then the input event is ``suppressed.''  \EATrans can
also function as an event mapper: that is when the output sequence
\anEventSeq is a single symbol \anotherEvent, meaning that \EATrans
replaces the input event \anEvent by an output event
\anotherEvent. Lastly, \EATrans may also inject events: that is when
\anEventSeq has multiple events.

In the rest of this paper, we may draw transition diagrams to
represent an edit automaton (Fig. ~\ref{fig-EA-1} and
\ref{fig-EA-2}). When we do so, the following convention is used.
\begin{convention} \label{conv-EA}
 The following conventions are followed when the transition diagram of
 an edit automaton is drawn.
 \begin{inparaenum}
 \item The state labelled \EAInit is the initial state.
 \item If the label ``$\anEvent \rightarrow \anEventSeq$'' appears on
   a transition edge from state \anEAState to $\anEAState'$, then that
   transition will be triggered on input event \anEvent, and the
   latter is transformed into the output event sequence $\anEventSeq$
   (i.e.,
   $\EATrans(\anEAState, \anEvent) = (\anEAState', \anEventSeq)$).
 \item We write ``\anEvent'' as a shorthand for the transition label
   ``$\anEvent \rightarrow \anEvent$'' (i.e., \anEvent is preserved).
 \item We write ``$\neg \anEvent$'' as a shorthand for the
   transition label ``$\anEvent \rightarrow \epsilon$'' (i.e., \anEvent is
   suppressed).
 \item If the labels ``$* \rightarrow \anEventSeq$'', ``$*$'', or
   ``$\neg *$'' appear on a transition edge from \anEAState to
   $\anEAState'$, then they mean
   ``$\anEvent \rightarrow \anEventSeq$'', ``\anEvent'', and
   ``$\neg \anEvent$'' respectively for every event
   $\anEvent \in \EASymbols$ that does not appear on any transition
   edge emanating from \anEAState.
 \item If none of the labels ``$* \rightarrow \anEventSeq$'', ``$*$'',
   or ``$\neg *$'' appears on any transition edge emanating from
   \anEAState, then implicitly there is a transition edge from
   \anEAState to \anEAState itself, with the label ``$\neg *$''.
 \end{inparaenum}
\end{convention}

\subsection{Case Study}
\label{sec-case-study}

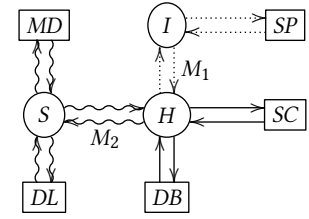
\begin{figure}
\centering
\mbox{
\xymatrix@R=17pt@C=28pt{
  *+[F]{\mathit{MD}} \ar@<.7ex>@{~>}[d]
  & *++[o][F]{I} \ar@<.7ex>@{..>}[d]^{$M_1$}
  \ar@<.7ex>@{..>}[r] &
  *+[F]{\mathit{SP}} \ar@<.7ex>@{..>}[l] \\
   *++[o][F]{S} \ar@<.7ex>@{~>}[r] \ar@<.7ex>@{~>}[u]
  \ar@<.7ex>@{~>}[d]
   &  *++[o][F]{H} 
     \ar@<.7ex>@{..>}[u] \ar@<.7ex>@{->}[d] 
     \ar@<.7ex>@{~>}[l]^{$M_2$}
     \ar@<.7ex>@{->}[r]
  & *+[F]{\mathit{SC}} \ar@<.7ex>@{->}[l] \\
  *+[F]{\mathit{DL}} \ar@<.7ex>@{~>}[u] &
  *+[F]{\mathit{DB}} \ar@<.7ex>@{->}[u] &
}
}
\caption{Brokers, devices, and network links in the smart home
  example. Ovals and boxes represent brokers and devices
  respectively. Dotted arrows represent links that are of type
  \textit{internet}; solid arrows, type \textit{door}; wavy arrows,
  type \textit{sensitive}. Edit automata $M_1$ and $M_2$ are deployed
  respectively along the link from $I$ to $H$ and the link from $H$ to
  $S$.\label{fig-smart-home-hierarchy}}
\end{figure}

We illustrate the use of our proposed architecture by an example smart
home configuration.

\paragraph{A Smart Home.}
The home owner has installed the following devices: a motion detector
(MD), a security camera (SC), a door lock (DL), and a door bell
(DB). In addition, the home owner carries a smart phone (SP).


Suppose all devices are connected to a centralized, cloud-based
broker.  For brevity, we identify an event by its topic in the
following discussion. The normal flow of events is listed below:
\begin{compactenum}[\ (1)]
\item MD is used for detecting if someone is currently at home.  MD
  periodically announces its findings by publishing either
  \texttt{MD\_motion} or \texttt{MD\_no\_motion}.
\item DB subscribes to \texttt{MD\_motion} and
  \texttt{MD\_no\_motion}, and updates its internal state accordingly.
\item When a person presses DB, the latter checks its internal state to
  see if anyone is at home. If someone is home, then the
  following steps are skipped.
\item If no one is at home, DB will request SC to take a picture of
  the person at the door by publishing \texttt{SC\_request}. Upon
  notification, SC will send a picture by publishing an
  \texttt{SC\_send} event with the picture as the payload.
\item Upon receiving the picture, DB publishes an event with 
  \texttt{AC\_request} as its topic and the picture as its payload.
\item SP subscribes to \texttt{AC\_request}. The home owner will
  examine the picture of the person at the door, and decide if access
  should be granted.  If so, then SP will publish
  \texttt{AC\_grant}. Otherwise, SP publishes \texttt{AC\_deny}.
\item When DB receives an \texttt{AC\_grant}, it will request DL to
  unlock the front door by publishing \texttt{DL\_unlock}.
\end{compactenum}

\begin{figure}
\centering
\mbox{
\xymatrix{
*+[o][F]{q_0}
\ar@(dr,ur)_{$*$, $\neg {\small\texttt{DL\_unlock}}$}
}
}
\caption{Edit automaton $M_1$ for enforcing Policy 3 (Visibility
  Control). $M_1$ has only one state, which suppresses
  \texttt{DL\_unlock} but preserves all other events.  Consult
  Convention \ref{conv-EA} for how transition diagrams are to be
  interpreted.\label{fig-EA-1}}
\end{figure}
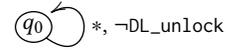

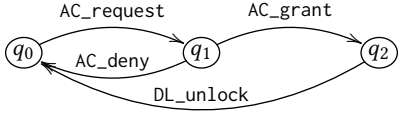
\begin{figure}
\centering
\mbox{
\xymatrix{
 *+[o][F]{q_0} 
 \ar@/^.8pc/[rr]^{\txt{\small\texttt{AC\_request}}}
 & &
 *+[o][F]{q_1} 
 \ar@/^.8pc/[rr]^{\txt{\small\texttt{AC\_grant}}}
 \ar@/^.8pc/[ll]_{\txt{{\small\texttt{AC\_deny}}}}
 & &
 *+[o][F]{q_2}
 \ar@/^1.8pc/[llll]_{\txt{\small\texttt{DL\_unlock}}}
}
}
\caption{Edit automaton $M_2$ for enforcing Policy 4 (Collaboration
  Protocol).  $M_2$ permits event sequences ``\texttt{AC\_request},
  \texttt{AC\_grant}, \texttt{DL\_unlock}'' and
  ``\texttt{AC\_request}, \texttt{AC\_deny}.'' Any extra
  \texttt{DL\_unlock} events that appear out of order are suppressed.
  Consult Convention \ref{conv-EA} for how transition diagrams are to
  be interpreted.\label{fig-EA-2} }
\end{figure}

\paragraph{Policies.} We desire to enforce four policies in the smart
home environment.
\begin{asparadesc}
\item[Policy 1 (Resiliency Against Cloud Disconnection).] If the cloud
  is ever disconnected, minimum services shall be available. For
  example, even though it is impossible to contact SP in the event of
  cloud disconnection, MD should still be able to update the internal
  state of DB.
\item[Policy 2 (Information Flow Control).] Events with topics
  \linebreak \texttt{MD\_motion} and \texttt{MD\_no\_motion} are
  sensitive information, and shall not be leaked outside of the home
  for privacy sake.
\item[Policy 3 (Visibility Control).] \texttt{DL\_unlock} is a
  sensitive control signal, and should only be sent from within the
  home. Any \texttt{DL\_unlock} event that is sent from the internet
  is discarded.
\item[Policy 4 (Collaboration Protocol).] It is possible that some of the devices (e.g., DB)
  are compromised. We want to make sure that \texttt{DL\_unlock} is
  published only as the last event in the sequence:
  \texttt{AC\_request}, \texttt{AC\_grant}, \texttt{DL\_unlock}. Any
  deviation from this collaboration protocol is a sign that some of
  the devices have been compromised.
\end{asparadesc}

To enforce the policies stated above, the smart home is
reconfigured according to Fig.~\ref{fig-smart-home-hierarchy}.
We explain the configuration below.

\paragraph{Enforcing Policy 1 (Resiliency Against Cloud
  Disconnection).}  Notice that multiple brokers are introduced
(Fig.~\ref{fig-smart-home-hierarchy}). The broker $I$ is hosted in the
cloud, which provides a point of contact to the SP.  The broker $H$ is
a gateway to the local network of the smart home.  Broker $S$ and
the home devices are connected to $H$ and to each other via a
LAN. If the cloud is ever disconnected (i.e., the network links
between $H$ and $I$ are severed), the rest of the broker ensemble will
still be functional.

\paragraph{Enforcing Policy 2 (Information Flow Control).}
As shown in Fig.~\ref{fig-smart-home-hierarchy}, the network links are
classified into three types: (a) \textit{sensitive}, (b)
\textit{door}, and (c) \textit{internet}.  A global brokering policy
is formulated in such a way that the following propagation is
\emph{not} allowed:
\begin{description}
\item[Brokering Policy.] Events received from a link of type \textit{sensitive} cannot
  be propagated to a link of type \textit{internet}.
\end{description}
All other forms of propagation are allowed. The overall effect of this
brokering policy is that \texttt{MD\_motion} and
\texttt{MD\_no\_motion} are not propagated by $H$ to $I$, thus
sensitive events remain inside the home.

\paragraph{Enforcing Policy 3 (Visibility Control).}
We want to filter away any \texttt{DL\_unlock} events that pass
through the network link from broker $I$ to broker $H$. This is
achieved by placing an edit automaton $M_1$ along that network link
(Fig.~\ref{fig-smart-home-hierarchy}). The definition of $M_1$ is
depicted in the transition diagram in Fig.~\ref{fig-EA-1}, which
preserves all events except for \texttt{DL\_unlock}.  Note two
points. First, Policy 3 cannot be enforced by brokering policies, as
we allow some events (e.g., \texttt{AC\_grant}, see below) from broker
$I$ to enter broker $S$, but disallow others (e.g.,
\texttt{DL\_unlock}) to do so.  Second, since $M_1$ has only one
state, Policy 3 does not exploit the history tracking feature of an
edit automaton.

\paragraph{Enforcing Policy 4 (Collaboration Protocol).}
Policy 4 is a history-based policy. Its enforcement requires the
stateful tracking of event history. To this end, an EA $M_2$ is
interposed along the network link from broker $H$ to broker $S$
(Fig.~\ref{fig-EA-2}).  $M_2$ permits the sequences
``\texttt{AC\_request}, \texttt{AC\_grant}, \texttt{DL\_unlock}'' and
``\texttt{AC\_request}, \texttt{AC\_deny}.'' Any out-of-order
occurrences of \texttt{DL\_unlock} are suppressed.  Notice that all
events with topics \texttt{AC\_request}, \texttt{AC\_grant},
\texttt{AC\_deny}, and \texttt{DL\_unlock} will pass through $H$,
which in turn propagates them to $S$.  In this way, the collaboration
protocol is enforced by $M_2$ before the events can reach DL.  This
protects the integrity of DL even if other devices are compromised.
Note that the interposing of $M_2$ between $H$ and $S$ requires the
system to track the automaton state.

In summary, we have demonstrated how the proposed protection
scheme supports resiliency against cloud disconnection, information
flow control, visibility control, and collaboration protocols.

\section{Operational Semantics}
\label{sec-model}

This section presents the operational semantics of our proposed
architecture in the form of a state machine model.  Minor variants of
this model have been ``mechanized'' using PLT Redex \cite{PLT-Redex},
a semantic engineering tool.  The PLT Redex encoding of our state
machine model consists of 820 lines of code, including both model
definition and test cases.  The mechanized model is executable.  This
exercise has allowed us to gain deeper insights into our formulation,
and to catch a few errors in our early work.  The mathematical
formulation and the PLT Redex encoding lay the semantic foundation for
future study of policy enforceability (see \S \ref{sec-conclusion}).

\subsection{System Configuration}

The configuration of an event-based system is captured in a
\Dfn{(system) schema} \aSystem, which is a quadruple \System{
  \aConGraph }{ \aBrokerPolicy }{ \anEventPolicy }{ \Sub }{ \TaskOrder
} with the components below:
\begin{compactitem}[$\bullet$]
\item \aConGraph is a \Dfn{connection graph} of the form \ConGraph{
    \Devices }{ \Brokers }{ \Link }:
\begin{itemize}
\item \Devices and \Brokers are two disjoint, finite sets of
  \Dfn{entities}. \Devices is the set of \Dfn{devices}, and \Brokers
  is the set of \Dfn{brokers}. We write \Entities to
  denote $\Devices \cup \Brokers$.
\item $\Link \subseteq \Entities \times \Entities$ is a binary
  relation over entities. It represents network connections.
  The binary relation \Link satisfies two additional requirements:
  (a) \Link is symmetric but irreflexive; (b) $\Link \cap (\Devices
  \times \Devices) = \emptyset$.
\end{itemize}
%
%
Furthermore, \Link
induces four binary relations:
(a) $\Publish = \Link \cap ( \Devices \times \Brokers )$ captures
  device-to-broker links, 
(b) $\Notify = \Link \cap ( \Brokers \times \Devices )$ captures broker-to-device links,
(c) $\Bridge = \Link \cap (\Brokers \times \Brokers)$ captures
broker-to-broker links, and (d) $\Monitored = \Link \setminus \Notify$
captures links with a broker as the destination.
\item \aBrokerPolicy is the \Dfn{brokering policy}, which is a
  structure of the form \BrokerPolicy{ \BPTypes }{ \BPTyping }{
    \BPAllow }:
\begin{itemize}
\item \BPTypes is a finite set of \Dfn{link types}. 
\item $\BPTyping : \Link \rightarrow
  \BPTypes$ assigns a link type to each 
  link. 
\item $\BPAllow \subseteq \BPTypes \times \BPTypes$ is a binary
  relation defined over \BPTypes. 
  If $\BPAllow ( \aBPType_1, \aBPType_2 )$, then a broker is allowed
  to pass along an event it receives from a link of type
  $\aBPType_1$ to a link of type $\aBPType_2$. 
\end{itemize}
    The brokering policy \aBrokerPolicy induces a ternary relation
    $\BPPropagate \subseteq \Entities \times \Entities \times
    \Entities$, so that $\BPPropagate ( \anEntity, \anotherEntity, \yetAnotherEntity)$
    iff $\Link (\anEntity, \anotherEntity)$, $\Link (\anotherEntity,
    \yetAnotherEntity)$, and $\BPAllow ( \BPTyping( \anEntity,
    \anotherEntity),
    \BPTyping (\anotherEntity, \yetAnotherEntity))$.  That is, 
  $\BPPropagate ( \anEntity, \anotherEntity, \yetAnotherEntity)$
    asserts that an event passing through link $(\anEntity,
    \anotherEntity)$ is allowed to be further propagated by the broker \anotherEntity
    through the link $(\anotherEntity, \yetAnotherEntity)$.
\item The \Dfn{event policy} \anEventPolicy is a quadruple
  \EventPolicy{ \Events }{ \EAStates }{ \EAInit }{ \Trans }:
\begin{itemize}
\item \Events, \EAStates, and \EAInit are the components of an EA.
  \Events is the set of \Dfn{events} (more precisely \Dfn{event
    topics}) that can be transmitted in the event-based system.
\item $\Trans : \Monitored \rightarrow 
   (\EAStates \times \Events \rightarrow 
     \EAStates \times \Events^*)$ assigns 
    an EA transition function
   to each monitored link.
\end{itemize}
  More specifically, the EA $\anEA( \anEntity,
  \anotherEntity ) = \EA{ \Events }{ \EAStates }{ \EAInit }{ \Trans(
  \anEntity, \anotherEntity) }$ is the EA that transforms the events
   sent from \anEntity to \anotherEntity.
 \item $\Sub : \Notify \rightarrow \Powerset{ \Events }$ assigns a set
   of events to each broker-to-device link.  Intuitively, $\Sub (
   \anEntity, \anotherEntity )$ is the set of events subscribed by
   device \anotherEntity in broker \anEntity.
\item $\sqsubseteq$ is a partial ordering defined over the set of
    \Dfn{annotated tasks}.  Intuitively, the dynamics of the system is
    modelled as the generation and discharging of tasks.  These tasks
    are ``queued up'' in a work list within the system state for further
    processing.  A \Dfn{task} \aTask is defined via the following
    grammar.
    \begin{align*}
      \aTask ::= \mbox{} & \TaskTransmit{ \anEntity }{ \anotherEntity
      }{ \anEvent } \mid \TaskBroker{ \anEntity }{ \anotherEntity }{
        \anEventSeq }
    \end{align*}
    where $\anEntity, \anotherEntity \in \Entities$, $\anEvent \in
    \Events$, and $\anEventSeq \in \Events^*$.  We write \Tasks{
      \aSystem } for the set of all tasks defined for schema \aSystem.
    
    An annotated task is a construct of the form \Annotate{ \aTask }{
      \aTSGen }{ \aTSPub }, in which the task \aTask is annotated
    with two timestamps (i.e., natural numbers), (i) \aTSGen, the
    generation time of \aTask, and (ii) \aTSPub, the generation time
    of the event publication task from which \aTask is derived.  We
    write \AnnTasks{ \aSystem } to denote the set of all annotated
    tasks for schema \aSystem, and write \anAnnTask for a typical
    member of \AnnTasks{ \aSystem }.

    By imposing the partial ordering \TaskOrder over \AnnTasks{
      \aSystem } to indicate how tasks are prioritized, one can
    simulate different \Dfn{relative network speeds} (see
    \cite[\S 3.3.6]{Thesis} for examples).  In particular, if
    $\anAnnTask \TaskOrder \anAnnTask'$, then \anAnnTask will be
    processed before $\anAnnTask'$.  The annotation of tasks allows
    \TaskOrder to be formulated in terms of timestamps.

\end{compactitem}

\subsection{System States}

Given a schema \aSystem, a \Dfn{system state} \aSysState is a triple
\SysState{ \aTS }{ \SSST }{ \SSWL }:
\begin{compactitem}[$\bullet$]
\item The system state tracks a global clock, for which $\aTS \in \Nat$ is
  the current time. The clock is used within the model for 
  producing timestamps.
\item $\SSST : \Monitored \rightarrow \EAStates$ is a function
  assigning an EA state to each link that is monitored by an EA.  In
  particular, $\SSST ( \anEntity, \anotherEntity )$ is the current
  state of $\anEA ( \anEntity, \anotherEntity )$, the EA guarding link
  $( \anEntity, \anotherEntity )$.
\item The \Dfn{work list} $\SSWL \subseteq \AnnTasks{ \aSystem }$ is a
  finite set of annotated tasks. In the following, the predicate
  $\TaskSelect( \anAnnTask, \SSWL )$
  asserts that \anAnnTask is a minimal
  element in \SSWL according to \TaskOrder. Note that for a given
  \SSWL there may be multiple annotated tasks satisfying the
  \TaskSelect predicate.  As usual, nondeterminism is implied in such
  cases.
\end{compactitem}
Let \SysStates{ \aSystem } be the set of all system states as defined
above. We write \aSysState to denote a typical member of \SysStates{
  \aSystem }.  The \Dfn{initial state} of a system is
$\theInitSysState = \SysState{ 0 }{ \SSST_{\mathit{init}} }{ \emptyset
}$, where $\SSST_{\mathit{init}}( \anEntity, \anotherEntity ) =
\EAInit$ for every $( \anEntity, \anotherEntity ) \in \Monitored$.

\subsection{State Transition}

%
%

Given a schema \aSystem, we define a state transition relation
$\cdot \SysTransBinRel{ \aSystem } \cdot \subseteq \SysStates{
  \aSystem } \times \SysStates{ \aSystem }$. Intuitively,
$\aSysState \SysTransBinRel{ \aSystem } \aSysState'$ means that
$\aSysState'$ is a successor state of \aSysState.  The transition
relation is defined by the following set of transition rules, which
specify the condition under which
$\aSysState \SysTransBinRel{ \aSystem } \aSysState'$, where
$\aSysState = \SysState{ \aTS }{ \SSST }{ \SSWL }$, and
$\aSysState' = \SysState{ \aTS' }{ \SSST' }{ \SSWL' }$.  In the
following specification, we follow the convention that by default
$\aTS' = \aTS + 1$, $\SSST' = \SSST$ and $\SSWL' = \SSWL$, unless the
rules explicitly say otherwise.

\begin{compactitem}[$\bullet$]
\item \textbf{T-Publish}. \emph{Generate an event
    publication task.}
\begin{compactitem}[$\circ$]
\item \textbf{Precondition:} 
$\Publish ( \anEntity, \anotherEntity )$, and $\anEvent \in \Events$.
\item
\textbf{Effect:} $\SSWL' = \SSWL \cup \{\, \Annotate{ \TaskTransmit{ \anEntity }{
    \anotherEntity }{ \anEvent } }{ \aTS }{ \aTS } \,\}$.
\end{compactitem}

\item \textbf{T-Notify}. \emph{Consume an event notification task.}
\begin{compactitem}[$\circ$]
\item
\textbf{Precondition:} $\TaskSelect(\anAnnTask,
   \SSWL)$, $\anAnnTask = \Annotate{ \aTask }{ \aTSGen }{
  \aTSPub }$, $\aTask = \linebreak \TaskTransmit{ \anEntity }{ 
 \anotherEntity }{
  \anEvent }$, and $\Notify ( \anEntity, \anotherEntity )$.
\item
\textbf{Effect:} $\SSWL' = \SSWL \setminus \{\, \anAnnTask \,\}$.
\end{compactitem}

\item \textbf{T-Deliver}. \emph{Transmit an event over a link, and
    apply execution monitor to the transmitted event.}
\begin{compactitem}[$\circ$]
\item
\textbf{Precondition:} $\TaskSelect(\anAnnTask, \SSWL)$,
  $\anAnnTask = \Annotate{ \aTask }{ \aTSGen }{ \aTSPub }$,
  $\aTask = \linebreak \TaskTransmit{ \anEntity }{ \anotherEntity }{
  \anEvent }$, and $\Monitored ( \anEntity, \anotherEntity )$.
\item
\textbf{Effect:}
 Let $\EATrans = \Trans ( \anEntity, \anotherEntity )$ and
  $( \anEAState, \anEventSeq ) = \EATrans (\SSST (\anEntity, \anotherEntity) , \anEvent)$.
  Then $\SSST' ( \anEntity, \anotherEntity ) = \anEAState$, and
 $\SSWL' =
  \SSWL_1 \cup \SSWL_2$, where:  
\begin{align*}
  \SSWL_1 = \mbox{} & \SSWL \setminus \{\, \anAnnTask \,\}\\
  \SSWL_2 = \mbox{} & 
     \begin{cases}
       \{\, \Annotate{ \TaskBroker{ \anEntity }{ \anotherEntity }{
            \anEventSeq } }{ \aTS }{ \aTSPub } \,\} & \text{if $\anEventSeq
             \neq \epsilon$}\\
        \emptyset & \text{otherwise}
     \end{cases}
\end{align*}


\end{compactitem}

\item \textbf{T-Broker}. \emph{Process a sequence of received events to create further transmissions.}
\begin{compactitem}[$\circ$]
\item
\textbf{Precondition:} $\TaskSelect(\anAnnTask, \SSWL)$, 
 $\anAnnTask = \Annotate{ \aTask }{ \aTSGen }{ \aTSPub }$,
 and $\aTask = \linebreak \TaskBroker{ \anEntity }{ \anotherEntity }{
  \anEvent \anEventSeq }$,
such that $\anEvent \in \Events$ and
$\anEventSeq \in \Events^*$.
\item
\textbf{Effect:} 
Let $Z = \{
   \yetAnotherEntity \in \Entities \mid 
   \Bridge ( \anotherEntity, \yetAnotherEntity ) \lor
   (\Notify ( \anotherEntity, \yetAnotherEntity ) \land
     \anEvent \in \Sub ( \anotherEntity, \yetAnotherEntity )
   )
\}$.
Then $\SSWL' = \SSWL_1 \cup \SSWL_2 \cup \SSWL_3$, where:
\begin{align*}
\SSWL_1 = \mbox{} & \SSWL \setminus \{\, \anAnnTask \,\}\\
\SSWL_2 = \mbox{} & 
  \begin{cases}
    \{\, \Annotate{ \TaskBroker{ \anEntity }{ \anotherEntity }{
       \anEventSeq } }{ \aTS }{ \aTSPub } \,\} & \text{if $\anEventSeq
     \neq \epsilon$}\\
   \emptyset  & \text{otherwise}
  \end{cases} \\
\SSWL_3 = \mbox{} & \{\, \Annotate{ \TaskTransmit{ \anotherEntity }{
  \yetAnotherEntity }{ \anEvent } }{ \aTS }{ \aTSPub } \mid \\ 
 & \qquad 
   \yetAnotherEntity \neq \anEntity \land
   \yetAnotherEntity \in Z \land 
   \BPPropagate (\anEntity, \anotherEntity, \yetAnotherEntity) \,\}
\end{align*}
\end{compactitem}
\end{compactitem}

\section{Usage Patterns}
\label{sec-usage}

We explore how the two protection mechanisms, brokering control and
execution monitoring, can be leveraged to impose various forms of
access control policies in an event-based system.

\subsection{Access Control with Brokering Policies}

The most liberal brokering policy is $\aBrokerPolicy_0 =
\BrokerPolicy{ \BPTypes_0 }{ \BPTyping_0 }{ \BPAllow_0 }$, where
$\BPTypes_0$ is a singleton set $\{ \aBPType_0 \}$, $\BPTyping_0$ maps
every link to $\aBPType_0$, and $\BPAllow_0 = \{ ( \aBPType_0,
\aBPType_0 ) \}$. This trivial brokering policy allows every event
received from a link to be forwarded to another link.  This set-up is
essentially the bridge feature of Mosquitto \cite{Mosquitto}: a bridge
connects two MQTT brokers, so that the events received by one broker
are made visible to the other broker.  This liberal brokering policy,
however, suffers from the following shortcoming. The brokers connected
by bridges form a single scope of events. There is no regulation of
what events are visible to which subscribers, making it difficulty to
confine the visibility of sensitive events (e.g., personal health
alerts).  
Imposing brokering policies more
restrictive than $\aBrokerPolicy_0$ allows us to regulate the flow of
information, as we illustrate in the following.

\subsubsection{Information Flow Control} \label{sec-BLP}

Brokering policies allow us to impose a form of information flow
control in the style of the Bell-LaPadula (BLP) model \cite{BLP}.  The
basic idea of the BLP scheme is that information sources (e.g.,
publishers) and information consumers (e.g., subscribers) are labelled
with security labels (e.g., unclassified, confidential, secret, top
secret). These security labels form a lattice structure, and is thus
partially ordered. BLP is essentially an access control model that
forbids ``reading up'' and ``writing down'' \cite{Rushby}.
Information may only flow from ``low'' information sources to ``high''
information consumers.

In our context, an event publication is a ``write,'' and an event
subscription can be considered a ``read.''  Rather than assigning
security labels to entities, we opt for a more uniform and flexible
approach, and assign security labels to links.  The key idea is that
when a broker \anotherEntity receives an event from a link
$(\anEntity, \anotherEntity)$, \anotherEntity is allowed to propagate
the event to a subsequent link $(\anotherEntity, \yetAnotherEntity)$
if the security label of the second link is at least as high as the
security label of the first link.  Consequently, the semantics of
assigning a security label \aBPType to a link is that events flowing
through that link comes from sources with security labels lower than
or equal to \aBPType.  When an event is transmitted through the
system, it goes through links with monotonically increasing labels
$\aBPType_1 \leq \aBPType_2 \leq \aBPType_3 \leq \ldots$.  More
specifically, one can configure the brokering policy $\aBrokerPolicy =
\BrokerPolicy{ \BPTypes }{ \BPTyping }{ \BPAllow }$ as follows to
control the flow of information within the system.
\begin{compactitem}[$\bullet$]
\item
  Let $(\BPTypes, \leq)$ be a partially ordered set of security labels.
\item The function \BPTyping assigns a security label to each link, in
  such a way that $\BPTyping(\anEntity, \anotherEntity) \leq
  \BPTyping(\anotherEntity, \yetAnotherEntity)$ whenever
  $\anotherEntity \in \Devices$.  That is, a notification link of a
  device \anotherEntity must have a security label no higher than the
  security label of every publication link of \anotherEntity, In other
  words, the device is ``reading down'' through the notification link,
  and ``writing up'' through the publication link.
\item Define \BPAllow so that $\BPAllow( \aBPType_1, \aBPType_2 )$ 
  only if $\aBPType_1 \leq \aBPType_2$.  That is, an event flowing
  through a link $(\anEntity, \anotherEntity)$ with a security label
  $l$ will only be propagated to a link $(\anotherEntity,
  \yetAnotherEntity)$ with a security label $h$ at least as high as
  $l$.
\end{compactitem}
With the scheme above, successive links that transmit an event will
have monotonically increasing security labels.  We formalize this
observation as follows.  

A sequence $\anEntity_0 \anEntity_1 \ldots \anEntity_n$ of entities in
\aConGraph is called a \Dfn{flow path} if (a)
$(\anEntity_i, \anEntity_{i+1}) \in \Link$ for $0 \leq i < n$, (b) for
$0 < i < n$, if $\anEntity_i \in \Brokers$, then
$\anEntity_{i-1} \neq \anEntity_{i+1}$ and
$\BPPropagate ( \anEntity_{i-1}, \anEntity_{i}, \anEntity_{i+1} )$.  A
flow path is a \Dfn{flow route} when none of the entities other than
the two ends (i.e., $\anEntity_1$, $\anEntity_2$, \ldots,
$\anEntity_{n-1}$) is a device.  A flow path is a potential path of
information flow through the system.  Intermediary entities along a
flow path can be devices which read a message and then propagate
information by publishing a correlated message.  A flow route is a
flow path for which the intermediary entities are all brokers.

With the way \BPAllow is defined, $\BPTyping ( \anEntity_{i-1},
\anEntity_{i} ) \leq \BPTyping ( \anEntity_{i}, \anEntity_{i+1} )$
when a broker $\anEntity_i$ relays a message from link
$(\anEntity_{i-1}, \anEntity_i)$ to link $(\anEntity_i,
\anEntity_{i+1})$.  Thus a message passes through links of
monotonically increasing security labels as it travels through a flow
route. A flow path is essentially the concatenation of flow routes for
which the concatenation points are devices.  The definition of
\BPTyping ensures that monotonicity is preserved by such
concatenation.

The administrator of the system may have some preconceived ideas about
what flow paths (resp.~routes) are permitted.  An important validation
task is to ensure that the configuration of the connection graph and
the brokering policy does not violate her expectation.  Given a
connection graph
$\aConGraph = \ConGraph{ \Devices }{ \Brokers }{ \Link }$, one can use
a variant of the Floyd-Warshall algorithm \cite[\S 26.2]{Algo} to
compute whether there is a legitimate flow path between each pair of
entities (more precisely, between each pair of links).  Described in
\cite[Appendix A]{Thesis}, the algorithm runs in $O(M^3)$ time,
where $M = |\Link|$.  The algorithm can be adopted to identify either
flow paths or flow routes.  Such an analysis allows us to debug the
topology of the connection graph and the assignment of security
labels, so as to ensure that devices that are supposed to communicate
with one another can do so, and flow paths that are not supposed to
exist are not accidentally enabled.


\subsubsection{Component Architecture} \label{sec-component}

The modular event-based systems proposed in \cite{Fiege-etal:2002:SAC,
  Fiege-etal:2002:KRR} is the idea that brokers are organized into a
hierarchy, so that each broker ``glues'' its children entities
together to form a distributed software component.  The Fiege \emph{et
  al.}  Scoping Rule of \S \ref{sec-background} provides a principled
way for these hierarchically organized software components to
communicate events with one another. We demonstrate in the following
that such a scoping rule can be simulated using brokering policies.

We assume that the hierarchy is specified through a parenthood
relation, $\mathit{parent} \subseteq \Entities \times \Entities$.
Intuitively, $\mathit{parent}(\anEntity, \anotherEntity)$ asserts that
\anotherEntity is a parent of \anEntity.  Two further restrictions
apply to the specification of $\mathit{parent}$.  First, a device is
never a parent of any entity. Second, parenthood chains never form a
cycle, not even a loop (i.e., a loop arises when an entity is its own
parent).

Based on the given parenthood relation, the following brokering policy
is formulated:
\begin{compactitem}[$\bullet$]
\item Let $\BPTypes = \{\, \mathsf{up}, \mathsf{down} \,\}$.
\item The function \BPTyping encodes the parenthood relation:
  $\BPTyping ( \anEntity, \anotherEntity ) = \mathsf{up}$
  if $\mathit{parent}( \anEntity, \anotherEntity )$, and
  $\BPTyping ( \anEntity, \anotherEntity ) = \mathsf{down}$ otherwise.
\item Define \BPAllow so that $\BPAllow( \aBPType_1, \aBPType_2 )$ holds
  unless $\aBPType_1 = \mathsf{down}$ and $\aBPType_2 = \mathsf{up}$.
  Effectively, a message can move ``up'' the hierarchy, and then
  ``down'', but never move ``up'' after it has moved ``down.''
\end{compactitem}
The brokering policy above ensures that events published in a broker
\anEntity is visible to subscribers in another broker \anotherEntity
if and only if \anEntity and \anotherEntity share a common ancestor in
the hierarchy.  To see this, observe that a flow route from one device
to another will only go from ``up'' to ``down'' but not vice versa.
This essentially means that the flow route will first reach a common
ancestor of the two devices before reaching its destination.  Thus the
sharing of a common ancestor determines visibility.

%
%

\subsection{Access Control with Edit Automata}
\label{sec-acc-con-EA}

We have already seen in the case study of \S \ref{sec-case-study} that
the history tracking feature of an edit automaton can be used for
enforcing collaboration protocols (e.g., Policy 4) among a group of
interacting devices.  We do not further elaborate on this use.

Event filtering and event mapping are popular visibility control
mechanisms in event-based systems. They are featured in the model of
Fiege \emph{et al.} \cite{Fiege-etal:2002:ECOOP, Fiege-etal:2002:KRR},
and implemented in Mosquitto \cite{Mosquitto}. We demonstrate that
these two features are special cases of execution monitoring by edit
automata. We then show how brokering policies can be combined with
edit automata to simulate complex protection domains in event-based
systems.

\subsubsection{Event Filtering}
Some events are used for coordination logics that belong to the
internal working of a distributed software component.  Publications of
such events should not be visible outside of the component because of
confidentiality considerations, and subscriptions of these events
should not extend beyond the component boundary in order that external
events do not affect the integrity of the component state.

Suppose $E \subseteq \Events$ is a subset of events that are allowed
to pass through a link.  Define the \Dfn{event filtering transition
  function} \Filter{E} so that $\EATrans ( \EAInit, \anEvent ) = (
\EAInit, \anEvent )$ if $\anEvent \in E$, and $\EATrans ( \EAInit,
\anEvent ) = ( \EAInit, \epsilon )$ if $\anEvent \not\in E$.  By
setting $\Trans( \anEntity, \anotherEntity ) = \Filter{E}$), only
events in $E$ will be transmitted along the link.

Event filtering leads naturally to the notion of event ``import'' and
``export'' for component architectures such as the one presented in \S
\ref{sec-component}.  A broker \anEntity can present an
\Dfn{interface} to each parent broker \anotherEntity. The interface
consists of a set $E_{\mathit{import}}$ of events that \anEntity is
willing to import from \anotherEntity, and a set $E_{\mathit{export}}$
of events \anEntity is willing to export to \anotherEntity.
To achieve the above, we set
$\Trans(\anEntity, \anotherEntity) = \Filter{ E_{\mathit{export}} }$,
and
$\Trans(\anotherEntity, \anEntity) = \Filter{ E_{\mathit{import}} }$.

\subsubsection{Event Mapping}

Event mapping is the transformation of events from one naming scheme
to another naming scheme when they are transmitted along a link.
Event mapping presents an alternative interface of a distributed
software component to a client component, often for ``gluing''
purposes or for information hiding.  Suppose
$f : \Events \rightarrow \Events$ is a function that renames events.
Define the \Dfn{event mapping transition function}
$\Mapper{f} : \EAStates \times \Events \rightarrow \EAStates \times
\Events^*$ such that
$\Mapper{f}(\EAInit, \anEvent) = (\EAInit, f(\anEvent))$.  That is,
the EA remains in state \EAInit at all time, and output $f(\anEvent)$
when the input event is \anEvent.  Setting
$\Trans(\anEntity, \anotherEntity)$ to \Mapper{f} will cause all
events passing through link $(\anEntity, \anotherEntity)$ to be
renamed according to $f$.

\subsubsection{RBAC-style Protection Domains}
\label{sec-RBAC}

In \cite{rbac-middleware}, a combination of publish/subscribe
middleware and Role-Based Access Control (RBAC) \cite{RBAC} is
employed to control the events that clients (i.e., devices) are
allowed publish and subscribe.  The idea is that clients are assigned
to roles, and membership in a role places limits on what events a
client can publish, as well as what events it is allowed to subscribe
to.  Although the language of RBAC is used, a ``role'' here is
essentially a protection domain in the event-based system.  (We are
not sanctioning RBAC for IoT!)  We describe in the following how this
model can be simulated by multiple brokers, brokering policies, and
execution monitors.

We assume that the following RBAC policy is given:
\begin{inparaenum}
\item a set \Devices of devices,
\item a set $R$ of roles,
\item a user-role assignment $\mathit{UA} \subseteq \Devices \times
  R$ which specifies role membership,
\item a function $\mathit{pub} : R \rightarrow \Powerset{\Events}$
  specifying for each role $r \in R$ the set $\mathit{pub}(r)$ of
  events that members of $r$ can publish, and
\item a function $\mathit{sub} : R \rightarrow \Powerset{ \Events }$
  specifying for each role $r \in R$ the set $\mathit{sub}(r)$ of
  events that members of $r$ can subscribe to.
\end{inparaenum}
Based on the parameters above, we describe how one can construct a
connection graph, a brokering policy, and edit automata to enforce the
RBAC policy.

First, we introduce a set of brokers:
\begin{align*}
  \Brokers & = \Brokers^{\mathit{roles}} \cup \{ \mathsf{Bus} \}
  &
    \Brokers^{\mathit{roles}} &
    = \Brokers^{\mathit{pub}} \cup \Brokers^{\mathit{sub}} \\
  \Brokers^{\mathit{pub}} & = \{ r^{\mathit{pub}} \mid r \in R \} &
  \Brokers^{\mathit{sub}} & = \{ r^{\mathit{sub}} \mid r \in R \}
\end{align*}
In short, we introduce two brokers, $r^{\mathit{pub}}$ and
$r^{\mathit{sub}}$, for each role $r$.  The intention is that clients
belonging to role $r$ will publish events to the \Dfn{publication
  broker} $r^{\mathit{pub}}$ and subscribes to events through
\Dfn{subscription broker} $r^{\mathit{sub}}$.  A distinguished broker
$\mathsf{Bus}$ is also introduced to perform message relaying among
the publication and subscription brokers.

Second, the links between the entities are defined as follows:
\begin{align*}
  \Link = \mbox{} & \Link_0 \cup \Link_0^{-1}\\
  \Link_0 = \mbox{} & \{ ( \anEntity, \mathsf{Bus} ) \mid \anEntity \in
                    \Brokers^{\mathit{roles}} \} \cup \mbox{} \\
   & \{ (\anEntity, r^{\mathit{pub}}) \mid \anEntity \in \Devices,
     (\anEntity, r) \in \mathit{UA} \} \cup \mbox{} \\
   & \{ (\anEntity, r^{\mathit{sub}}) \mid \anEntity \in \Devices,
     (\anEntity, r) \in \mathit{UA} \}                    
\end{align*}
The link set \Link is the symmetric closure of $\Link_0$, in which all
publication and subscription brokers are connected to the broker
$\mathsf{Bus}$, and every device belonging to role $r$ is connected to
the brokers $r^{\mathit{pub}}$ and $r^{\mathit{sub}}$.

Third, the following brokering policy is defined:
\begin{align*}
  \BPTypes = & \{\, \mathsf{SubUp}, \mathsf{SubDown},\\
    & \qquad \mathsf{PubUp},
               \mathsf{PubDown}, 
          \mathsf{DevUp}, \mathsf{DevDown}
               \,\} \\
  \BPTyping(\anEntity, \anotherEntity) =
   & \begin{cases}
      \mathsf{SubUp} & \text{if $\anEntity \in
        \Brokers^{\mathit{sub}}$ and $\anotherEntity =
        \mathsf{Bus}$}\\
      \mathsf{SubDown} & \text{if $\anEntity = \mathsf{Bus}$ and
        $\anotherEntity \in \Brokers^{\mathit{sub}}$} \\
      \mathsf{PubUp} & \text{if $\anEntity \in
        \Brokers^{\mathit{pub}}$ and $\anotherEntity =
        \mathsf{Bus}$}\\
      \mathsf{PubDown} & \text{if $\anEntity = \mathsf{Bus}$ and
        $\anotherEntity \in \Brokers^{\mathit{pub}}$}\\
      \mathsf{DevUp} & \text{if $\anEntity \in
        \Devices$ and $\anotherEntity \in
        \Brokers^{\mathit{roles}}$}\\
      \mathsf{DevDown} & \text{if $\anEntity \in \Brokers^{\mathit{roles}}$ and
        $\anotherEntity \in \Devices$}
    \end{cases}\\
  \BPAllow = & \{\, (\mathsf{PubUp}, \mathsf{SubDown}),\\
             & \qquad  (\mathsf{DevUp}, \mathsf{PubUp}),
               (\mathsf{SubDown}, \mathsf{DevDown}) \, \}
\end{align*}
In short, an event can only travel from a device to a publication
broker, then to broker $\mathsf{Bus}$, then to a subscription broker,
and lastly to another device.

Fourth, edit automata that perform event filtering are imposed along
every link from a publication broker to $\mathsf{Bus}$, and from
$\mathsf{Bus}$ to a subscription broker.
\[
  \Trans( \anEntity, \anotherEntity) =
  \begin{cases}
    \Filter{\mathit{sub}(r)} & \text{if $\anEntity = \mathsf{Bus}$ and
    $\anotherEntity = r^{\mathit{sub}}$}\\
    \Filter{\mathit{pub}(r)} & \text{if $\anEntity =
      r^{\mathit{pub}}$ and $\anotherEntity = \mathsf{Bus}$}
  \end{cases}
\]

While the given RBAC policy can be simulated by the above broker ensemble,
brokering policy, and execution monitors, two practical
considerations remain.

First, it is important to ensure that only the devices who are members
of role $r$ may connect to brokers $r^{\mathit{pub}}$ and
$r^{\mathit{sub}}$.  For example, the MQTT broker implementation
Mosquitto \cite{Mosquitto} allows the administrator to set up a local
ACL for each broker. Such an ACL can be used for encoding and
enforcing role membership.  More sophisticated access control
paradigm, such as ABAC, can be used for authorizing a device into a
role (i.e., a protection domain).

A second practical consideration is that, if a device belongs to
multiple roles, it may end up receiving the same event multiple
times. One way out is for the device to ``activate'' one role at a
time, or ``activate'' a subset of roles for which the subscribable
event sets are disjoint.  In other words, even if a device belongs to
multiple roles, it can connect to a subset of them in order to avoid
repeated delivery of the same message.

An extension of the scheme is to support role hierarchies (i.e.,
hierarchical nesting of protection domains).  The idea is that a
senior role (more capable protection domain) inherits all the
publishable events and subscribable events of its junior roles (less
capable protection domains). This can be simulated readily by
organizing the publication and subscription brokers hierarchically to
mirror the role hierarchy.

\section{Implementation}
\label{sec-implementation}

To evaluate the feasibility and the performance of brokering control
and execution monitoring for distributed event systems, we have
implemented a prototype of these two protection mechanisms in the
open-source Mosquitto broker \cite{Mosquitto, mosquittopaper} for the
MQTT protocol \cite{MQTT}.  MQTT is a messaging transport protocol
that makes use of the Publisher-Subscriber design pattern to provide
one-to-many message distribution and decoupling features
\cite{buschmann}.  It is used (or supported) widely in existing IoT
frameworks, including the IBM Watson IoT platform, the Amazon AWS IoT
platform, and the Microsoft Azure IoT platform.

%
%

Our implementation is based on the source code of Mosquitto version
1.4.10, the latest version when this work started.  The ``diff''
between the original Mosquitto code base and our extension involves
1,426 lines of C code across 19 source files (including 7 new files we
introduced into the code base).

\subsection{Multiple Brokers}

We reused the bridge facility offered by Mosquitto. This feature
allows a broker to be connected to another broker, so that the first
(aka child broker) behaves like a regular MQTT client of the second
(aka parent broker).  Events received by the child broker will be
propagated to the parent broker, and vice versa.  The interconnection
is established by a directive in the configuration file of the child
broker.  To support event filtering, one can set up the configuration
file so as to limit what event topics can be propagated from the child
to the parent, and what other topics can be propagated in the other
direction.  A degenerate configuration is to allow the propagation of
all topics, in both directions.  Brokering policies and execution
monitors are implemented on top of this bridge facility.

\subsection{Brokering Policies}

We extended Mosquitto so that each broker can be configured with an
arbitrary brokering policy. Specifically, we extended the syntax of
the Mosquitto configuration file, so that the administrator can (a)
define an arbitrary number of link types, (b) specify the \BPAllow
relation by enumerating all the pairs $(\aBPType_1, \aBPType_2)$ that
are in \BPAllow (or alternatively, identify those that are not in
\BPAllow), and (c) assign a link type to each link that goes into or
out of the broker.  This configuration mechanism is more general than
the formal model presented in \S \ref{sec-model}, as each broker can
be configured separately.

The broker implementation is extended to honor the brokering policy
specified in the configuration file.  Specifically, when an event is
received by the broker, the original Mosquitto implementation (the
\texttt{\_subs\_process} function in \texttt{subs.c}) enqueues this
event to all intended recipients (including regular MQTT clients,
child brokers, and parent brokers).  We modified this step by not
enqueuing an event for a recipient in case that propagation is not
permitted by the brokering policy.

\subsection{Execution Monitors}

To accommodate user-defined execution monitors, we have implemented a
``plug-in'' architecture for administrator to specify arbitrary
execution monitors as shared libraries (\texttt{.so} files on Linux),
and to configure the broker to interpose these execution monitors
along subsets of links.

\subsubsection{Monitor Modules}

The developer of an execution monitor can implement an edit automaton
as a shared library in the C programming language.  We have designed
an Application Programming Interface (API) for facilitating the
interaction between the extended Mosquitto implementation and these
edit automata.  The execution monitor is exported to the broker as a
``vtable'', which is a table of function pointers, pointing to (a) the
constructor and destructor functions of an edit automaton instance
(recall that execution monitoring is stateful), and (b) the transition
function.

When the shared library is loaded via dynamic linking into the broker
at the time of start-up, the constructor function will be called to
initialize the automaton instance.  The destructor will be used for
cleaning up the automaton instance when the broker shuts down.
Whenever a new message is received by the broker along a link
monitored by the edit automaton instance, the transition function will
be invoked with three arguments: (a) the message, (b) the current
automaton state, and (c) an \Dfn{event queue}.  The transition
function has two responsibilities.  First, the transition function
updates the automaton state. Second, the transition function produces
a sequence of output events. The latter is achieved by enqueueing the
output events into the event queue.

To facilitate the implementation of the transition function, the
broker context is captured into a vtable, which, again, is a table of
function pointers, pointing to broker facilities that the transition
function may take advantage of. These broker facilities include, for
example, dynamic memory allocation and logging. More importantly, the
broker context vtable also provides facilities for the transition
function to enqueue messages into the message queue.  This is how the
transition function can implement suppression, transformation, and
injection of events.  Elaborate design and implementation work has
been performed to make sure that such a framework interoperates well
with the message flow of all the three Quality-of-Service (QoS) levels
of MQTT.  Interested readers are directed to \cite[\S 4.3.3]{Thesis}
for such details.

\subsubsection{Configuration} We have extended the syntax of the
Mosquitto configuration file so that the administrator can specify
which shared libraries will be loaded as execution monitors.  The
syntax also allows one to specify which subset of network links will
be monitored by a given execution monitor.  More specifically, we
defined four \Dfn{message flow directions}, each corresponding to a
subset of network links that can be monitored by an edit automaton:
\begin{inparaenum}
\item \textbf{Import Publication (im\_pub)}: an incoming link
  from a child entity (i.e., an MQTT client or a child
  broker);
\item \textbf{Import Subscription (im\_sub)}: an outgoing link to 
  a child entity;
\item \textbf{Export Publication (ex\_pub)}: an incoming link from a parent broker;
\item \textbf{Export Pubscription (ex\_sub)}: an outgoing link to a parent broker.
\end{inparaenum}

\section{Performance Evaluation}
\label{sec-performance}

An empirical study has been conducted to profile the impact that
multiple brokers, brokering policies, and execution monitoring have on
the performance of an event-based system. In similar studies
\cite{babovic2016web, gutwin2011real}, two standard measurements used
to compare performance in networking technologies are (a) latency and
(b) message throughput. Latency is the time it takes to deliver one
message from one designated point to another (e.g. from a publisher to
a subscriber), whereas message throughput is the rate in which
messages are delivered.

We evaluated the performance impact of the proposed protection
mechanisms by measuring how they affect message throughput.  This is
achieved by comparing the message throughput of the original Mosquitto
implementation against that of the extended version reported in \S
\ref{sec-implementation}.  We chose to measure and report message
throughput rather than latency due to the following reasons:
\begin{inparaenum}
\item Message throughput is commonly used to establish requirements
  and limits for IoT technologies \cite{awslimits, microsoftlimits}.
\item Preliminary experiments we conducted to measure latency showed
  negligible differences between the original version of Mosquitto
  and the extended version.
\item The aforementioned preliminary experiments were
  susceptible to latency peaks, which resulted in inconsistent
  results, sometimes even favouring the extended version of
  Mosquitto. 
\item In similar experiments conducted by Babovic \emph{et al.}
  \cite{babovic2016web}, they recognize the importance of measuring
  message throughput to minimize error.
\end{inparaenum}

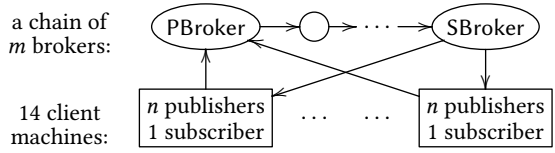
\begin{figure}
\centering
\mbox{%
\xymatrix@C=8pt@R=12pt{
  \txt{a chain of\\$m$ brokers:} &
  *++[o][F]{\mathsf{PBroker}} 
  \ar[r]
  & *++[o][F]{} \ar[r]
  & \cdots \ar[r] 
  & *++[o][F]{\mathsf{SBroker}}\\
  \txt{$14$ client\\machines:} &
  *+[F]{\txt{$n$ publishers\\
       1 subscriber}} \ar@{->}[u] \ar@{<-}[urrr]
   & \cdots & \cdots &
   *+[F]{\txt{$n$ publishers\\
       1 subscriber}} \ar@{->}[ulll] \ar@{<-}[u]
}%
}
\caption{Connections between brokers and client machines in
  experimental scenarios.\label{fig-expr-connections}}
\end{figure}

\subsection{Experimental Setup}

In our experiments, we compared the message throughput of five
\Dfn{configurations} (\S \ref{sec-configurations}) of Mosquitto
deployed in three different \Dfn{scenarios} (\S
\ref{sec-scenarios}). For each configuration-scenario combination, we
repeated the experiment for six \Dfn{rounds} (\S \ref{sec-rounds}).  A
different number of messages was published in each round, and we
measured the resulting message throughput in each case.

\begin{figure*}
  \centering
  \begin{subfigure}[b]{.3\textwidth}
    \centering
    \includegraphics[width=\textwidth]{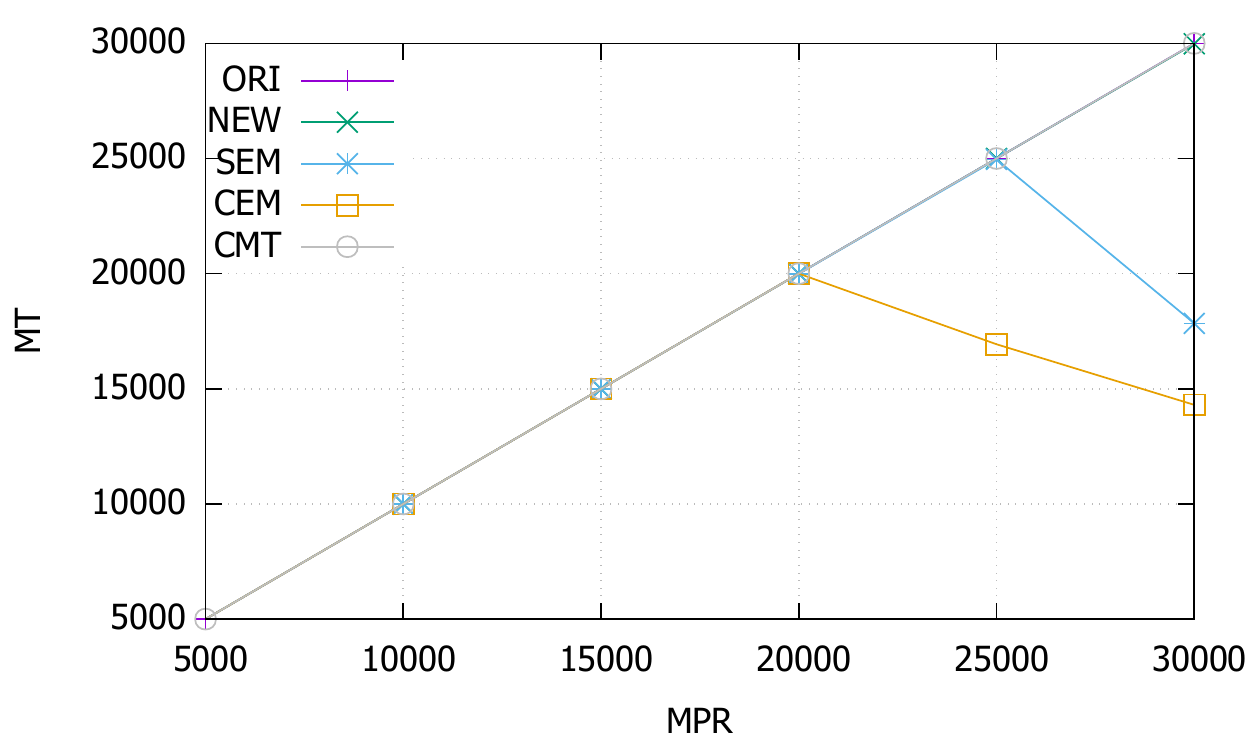}
    \caption{Scenario 1: with 1 broker\label{fig-results-1}}
  \end{subfigure}
  \hfill
  \begin{subfigure}[b]{.3\textwidth}
    \centering
    \includegraphics[width=\textwidth]{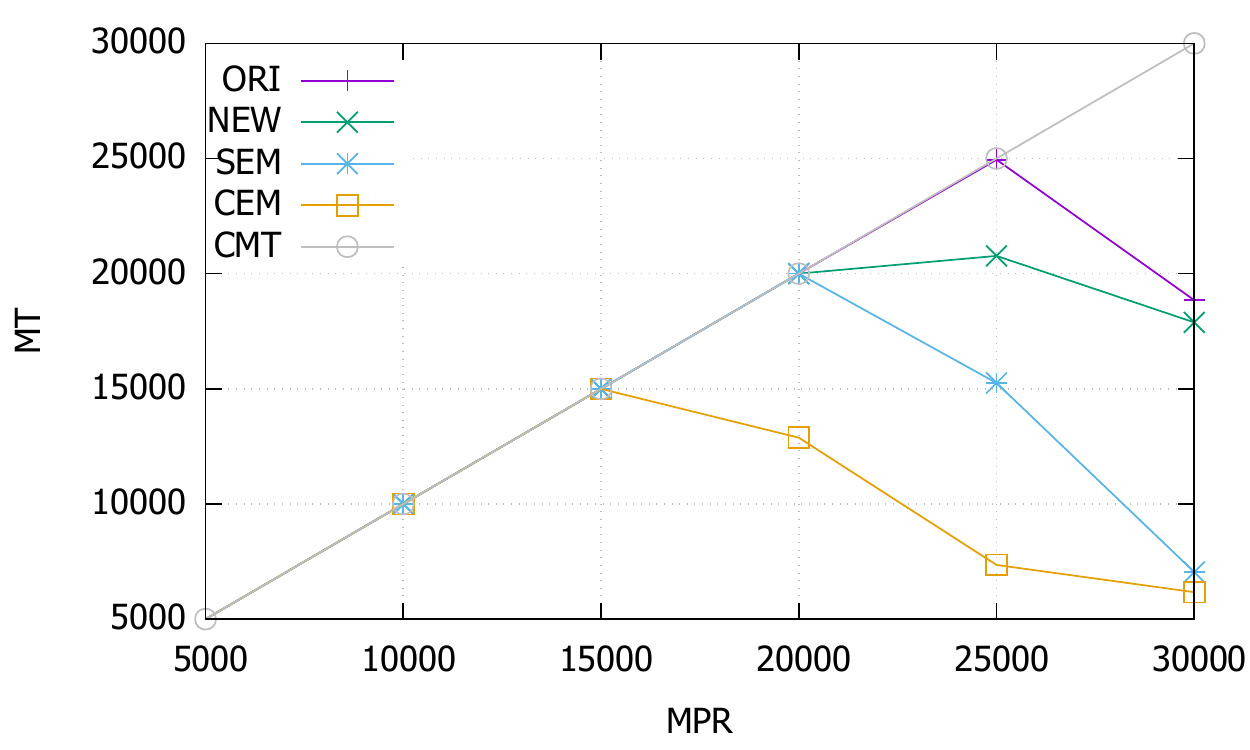}
    \caption{Scenario 2: with 3 brokers\label{fig-results-2}}
  \end{subfigure}
  \hfill
  \begin{subfigure}[b]{.3\textwidth}
    \centering
    \includegraphics[width=\textwidth]{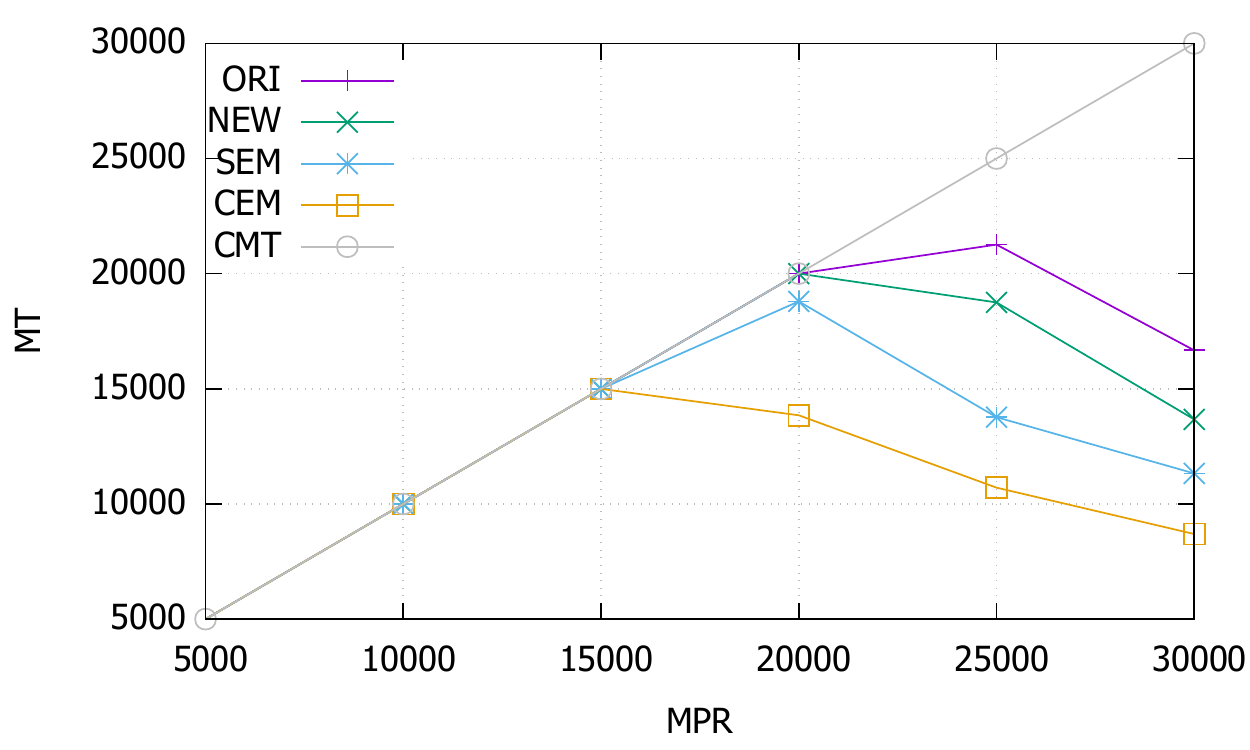}
    \caption{Scenario 3: with 5 brokers\label{fig-results-3}}
  \end{subfigure}
  \caption{Results of experiments: Message Throughput (MT) versus
    Message Publish Rate (MPR)\label{fig-results}}
\end{figure*}

\subsubsection{Broker Configurations}
\label{sec-configurations}

Two versions of Mosquitto were used in our study. We use
``\Dfn{original Mosquitto}'' to refer to Mosquitto 1.4.10, and
``\Dfn{extended Mosquitto}'' to refer to the implementation reported
in \S \ref{sec-implementation}.  We prepared five different broker
configurations.

\begin{asparaenum}
\item \textbf{ORI}. This configuration runs the original Mosquitto,
  and serves as the baseline for our experiments.  
\item \textbf{NEW}. This configuration runs the extended Mosquitto,
  but \emph{no} edit automaton is installed.  This configuration is for
  evaluating the performance impact of the execution monitor plugin
  infrastructure in extended Mosquitto.
\item \textbf{SEM}. This configuration runs extended Mosquitto with a
  simple execution monitor that passes on the input message unaltered.
  This is the fastest possible execution monitor.
\item \textbf{CEM}. This configuration runs extended Mosquitto with a
  complex execution monitor that does some random work for each byte
  in the message payload, and produces a random output message with
  the same size as the input message.  This typifies a
  monitor that runs in time proportional to the message
  size.
\item \textbf{CMT}.  This is not really an actual broker
  configuration, but a hypothetical, idealized broker configuration
  that incurs zero latency.  Its message throughput, which we call
  \Dfn{Computed Message Throughput (CMT)}, is always the same as the
  message publish rate.\footnote{In the other experiments we conducted
    to evaluate how message suppression and injection by edit automata
    affect message throughput, CMT takes a much more complex form. See
    \cite[\S 5.3.1]{Thesis} for more details.}  This is
  essentially a theoretical upper bound for the message throughput.
\end{asparaenum}

The installed edit automata in SEM and CEM process the messages
received by the broker, before the broker propagates the messages to
neighbouring brokers and subscribers.  If multiple brokers are deployed
in a scenario, then the same edit automaton is installed in every
broker.

Lastly, if a scenario involves multiple brokers (\S
\ref{sec-scenarios}), then ORI uses the bridge feature of Mosquitto,
and the rest of the configurations use the simple brokering policy
that implements Fiege \emph{et al.}'s scoping rule (see \S
\ref{sec-background} and \S \ref{sec-component}).

\subsubsection{Connection Scenarios}
\label{sec-scenarios}

We experimented with different number of interconnected brokers in
order to evaluate how multiple brokers and brokering policies
influence the message throughput.  Fig.~\ref{fig-expr-connections}
show how brokers and client machines are connected to one another in a
typical experimental scenario.  Scenarios differ by having a different
number of brokers.  In each of the three scenarios, there are $m$
brokers, and their connections form a chain.  The value of $m$ is $1$,
$3$, and $5$ respectively for Scenario 1, 2, and 3.

At the beginning of the chain is a broker that we designate
$\mathsf{PBroker}$, and the last broker in the chain is designated
$\mathsf{SBroker}$.  In Scenario 1, 
both $\mathsf{PBroker}$ and $\mathsf{SBroker}$ refer to the same broker.

Publishers and subscribers are spawned across fourteen different
client machines, where each machine launches a single subscriber and
$n$ publishers. The exact value of $n$ depends on the setup of the
round (see \S \ref{sec-rounds}).  Publishers publish to
$\mathsf{PBroker}$, and subscribers register their subscriptions on
$\mathsf{SBroker}$.

\subsubsection{Publishing Rounds}
\label{sec-rounds}

Multiple rounds of experiments are conducted for each 
configuration-scenario combination.  In each round, a fixed
number of events per second were published for a period of 10 minutes
(600 seconds).  We use the term \Dfn{Message Publish Rate (MPR)} to
refer to the speed in which messages are published in a given round.
MPR is measured in \Dfn{message per second (mps)}.  More specifically,
six rounds of messages were published for each configuration-scenario
combination, with an MPR of 5K, 10K, 15K, 20K, 25K, and 30K mps
respectively.  Since fourteen client machines are used for launching
publishers, a total of $m = \mathit{MPR}/14$ publishers were
spawned on each client machine.

Every published message has the same payload size of 175 bytes. This
size was chosen to represent the typical payload of sensor
readings. An example of an MQTT message carrying information about the
status of a battery is reported in \cite{mqttsensor}, with a payload
size of 97 bytes. A payload size of 175 bytes is therefore a
conservative overestimate.  The message topics and the subscriptions
are arranged so that every message is intended for exactly one
subscriber.

We then measured the overall Message Throughput (MT), which is also
measured in mps.  The MT of the publishing round is obtained by the
following formula:
\[
    \mathit{MT} = (\texttt{\# messages received by all
      subscribers}) \div 600 \mathit{sec}
\]

\subsubsection{Hardware and Software}

Fourteen machines were used to launch publishers and subscribers, and
(up to) five machines were used to run Mosquitto brokers.  These
machines are virtual machines hosted in the following hardware:
\begin{inparaitem}
\item Chasis: IBM BladeCenter H type
\item Storage: SAN = 600 GB + 300 GB + 2TB
\item Visualization Software: WMware ESXi, 4.1.0, 800380
\item Processor Type: Intel(R) Xeon(R) CPU X5660 @ 2.80GHz
\item CPU Cores: 12 CPUs x 2.8 GHz
\end{inparaitem}

All virtual machines have the following specifications:
\begin{inparaitem}
\item OS: Centos release 6.9 (Final)
\item CPU:  Intel(R) Xeon(R) CPU  X5660  @ 2.80GHz (1 core)
\item Memory: 8 GB
\item Disk Space: 14 GB
\end{inparaitem}

The following software are used in our experiments.
\begin{inparaenum}
\item \textbf{Mzbench \cite{mzbench}:} Load Testing Tool where users are able
  to write benchmarking scenarios (in Erlang or Python) for testing
  applications with different protocols.
\item \textbf{vmq\textunderscore mzbench \cite{vmq}:} Mzbench worker for the
  MQTT protocol. Publishers and subscribers used in the experiment are
  Mzbench scenarios which use the vmq\textunderscore mzbench worker.
\end{inparaenum}

\subsection{Results and Analysis}

The Message Throughputs (MT) of the various configurations is plotted
against different Message Publish Rates (MPR) in
Fig.~\ref{fig-results}, with one plot for each of the three scenarios.

Fig.~\ref{fig-results-1} shows the results for Scenario 1, which
involves only one broker.  This scenario highlights the relative
performance of the various configurations in the absence of
interconnected brokers.  Not only did the original Mosquitto
implementation (ORI) achieve the idealized Message Throughput (CMT),
the extended Mosquitto implementation (NEW) can do the same when no
execution monitor is installed.  This means that when the security
mechanisms are inactive, the performance impact of the added software
infrastructure is negligible.

In Fig.~\ref{fig-results-1}, we also notice that when edit automata
are installed, the Message Throughput began to drop after
the Message Publish Rate reached a certain point.  This is when the
broker was overwhelmed, and started to drop messages.  The best achieved
Message Throughput prior to message dropping signifies the capacity of
the configuration in question: 25K mps for SEM, and 20K mps for CEM.
This means that as the edit automaton does more work, the best
achieved Message Throughput degrades. This pattern persists in
Scenario 2 and 3 (Fig.~\ref{fig-results-2} and \ref{fig-results-3}).

Fig.~\ref{fig-results-2} and \ref{fig-results-3} show the impact of
multiple brokers and brokering policies.  Now even the ORI and NEW
configurations failed to match the idealized Message Throughput (CMT).
As expected, the more brokers are involved, the higher the latency,
and the best achieved Message Throughput will decline.  Scenario 3
introduces the most latency of the three scenarios, as five
interconnected brokers were involved.  In that case, the best achieved
Message Throughput of CEM was 15K mps.

To put our results into perspective, consider the throughput limits of
the IoT cloud services offered by Amazon and Microsoft.
\begin{compactitem}[$\bullet$]
\item Amazon established a limit of 9K publish requests per AWS
  account for their AWS Services (Table Message Broker Limits in the
  website) \cite{awslimits}. This means that each AWS account can
  publish a maximum of 9K mps to their Message Broker.
\item Microsoft divides their IoT Hubs solutions into three tiers: S1,
  S2 and S3 \cite{microsoftlimits}. Tier S3 is the most powerful
  of them, allowing an average message throughput of 208,333 messages
  per minute (i.e., approximately 3.5K mps).
\end{compactitem}
Comparing these values against best achieved Message Throughput of 15K
mps achieved by CEM (the configuration with the slowest execution
monitor among the five) in Scenario 3 (the scenario with the highest
latency among the three), our proposed protection scheme appears to
have no trouble coping with the throughput requirements expected in
commercial IoT frameworks.

Additional experiments are reported in \cite[\S 5.3.4]{Thesis}. These
experiments study how Message Throughput is impacted by edit automata
that either suppress or inject messages.

\section{Conclusion and Future Work}
\label{sec-conclusion}

In this work, we have demonstrated the utility and performance of
three protection mechanisms for event-based systems: multiple brokers,
brokering policies, and execution monitors.  A number of research
directions are listed here.
\begin{inparaenum}
\item We would like to be able to specify the overall collaboration
  protocol of an ensemble of devices, and then compile the
  specification to low-level brokering policies and edit automata for
  enforcement.  This relieves the user from having to write error-prone
  scripts, as is now practised in many commercial IoT frameworks.
\item Exactly what family of security policies are enforceable
  \cite{Schneider:2000, EA:IJIS, EA:TISSEC, Fong:2004} by the
  combination of multiple brokers, brokering policies, and execution
  monitors? We believe that the answer to this question depends on the
  relative network speed assumed in the formal operational semantics.
\item Using brokers as protection domains (\S \ref{sec-RBAC}) allows
  us to impose access control over the senders and receivers of
  messages. Yet one may want to also restrict the contents of the
  messages being sent.  How content-based access control can be
  imposed on messages passing through network links in a manner
  efficient enough to meet the throughput demand of IoT applications
  is a technical challenge.
\end{inparaenum}

\bibliographystyle{plain}

\end{document}